\documentclass[referee]{aa} 
\usepackage{graphicx}
\usepackage{natbib}

\begin{document}

\newcommand{\gapprox}{$\stackrel {>}{_{\sim}}$}   
\newcommand{\lapprox}{$\stackrel {<}{_{\sim}}$}

\title{Evidence for T Tauri-like emission in the EXor V1118 Ori from near-IR and X-ray data
\thanks{Based on observations collected at the AZT-24 telescope (Campo
 Imperatore, Italy)}}

\author{D.Lorenzetti
          \inst{1},
          T.Giannini
          \inst{1},
          L.Calzoletti
          \inst{1,2},
          S.Puccetti
          \inst{1,3},
          S.Antoniucci
          \inst{1,3},
          A.A.Arkharov
          \inst{4},
          A.Di Paola
          \inst{1},
          V.M.Larionov
          \inst{5}
          and
          B.Nisini
          \inst{1}
          }
\offprints{Dario Lorenzetti, email:dloren@mporzio.astro.it}

\institute{INAF - Osservatorio Astronomico di Roma, via Frascati
33, I-00040 Monte Porzio (Italy) \and Universit\`a degli Studi di
Cagliari - Dipartimento di Fisica, S.P. Monserrato-Sestu Km 0.700,
I-09042 Monserrato-CA (Italy) \and Universit\`a degli Studi di
Roma "Tor Vergata" - Dipartimento di Fisica, via della Ricerca
Scientifica 1, I-00133 Roma (Italy) \and Central Astronomical
Observatory of Pulkovo, Pulkovskoe shosse 65, 196140
St.Petersburg, Russia \and Astronomical Institute of St.Petersburg
University, Russia\\
\email{dloren,giannini,calzol,puccetti,antoniucci,dipaola,\\
nisini@mporzio.astro.it; arkharov@mail.ru; vml@VL1104.spb.edu} }
\date{Received;Accepted}

\abstract{}{}{}{}{}

  \abstract
   {}
   {We present a near-IR study of the EXor variable V1118 Ori,
performed by following a slightly declining phase after a recent
outburst. In particular, the near-IR (0.8 - 2.3 $\mu$m) spectrum,
obtained for the first time, shows a large variety of emission
features of the HI (Paschen and Brackett series), \ion{He}{i}
recombination and CO overtone.}
   {By comparing the observed spectrum with a wind model, a
mass loss rate of 4 10$^{-8}$ M$_{\sun}$ yr$^{-1}$ can be derived
along with other parameters whose values are typical of an
accreting T Tauri star. In addition, we have used X-ray data from
the XMM archive, taken in two different epochs during the
declining phase monitored in IR. X-ray emission (in the range 0.5
- 10 keV) permits to derive several parameters (as plasma
temperatures and L$_X$ luminosity) which confirm the T Tauri
nature of the source.}
   {In the near-IR the object maintains a low extinction
(A$_V$ \lapprox 2) during all the activity phases, confirming that
variable extinction does not contribute to brightness variations.
The lack of both a significant amount of circumstellar material
and any evidence of IR cooling from collimated jet/outflow driven
by the source, indicates that, at least this member of the EXor
class, is in a late stage of the Pre-Main Sequence evolution.
Going from inactive to active phases the luminosity increases
considerably (from 1.4 $L_{\sun}$ to more than 25 $L_{\sun}$) and
the observed spectral energy distribution (SED) assumes different
shapes, all typical of a T Tauri star. In the X-ray regime, an
evident fading is present, detected in the post-outburst phase,
that cannot be reconciled with the presence of any absorbing
material. This circumstance, combined with the persistence (in the
pre- and post-outburst phases) of a temperature component at about
10 MK, suggests that accretion has some influence in regulating
the coronal activity.}
   {}


\authorrunning{Lorenzetti D. et al.}
\titlerunning{Near-IR and X-ray studies of V1118 Ori}
\maketitle

%

\section{Introduction}

The star V1118 Ori is considered a pre-main sequence EXor variable
which undergoes subsequent outbursts in optical light. The EXors
class (whose name derives from the prototype EX Lup) is composed
by nine members defined and listed by Herbig (1989). These sources
are phenomenologically characterized by repetitive outbursts with
an amplitude of up to 5 magnitudes, lasting one year or less, with
a recurrence time of about 5-10 years. They are supposed to live
an intermediate phase between the strong FUor eruptions and the
more quiescent main sequence (Herbig 1977, Hartmann, Kenyon \&
Hartigan 1993), during which they exhibit an activity level lower
than the FUors one. The differences between FUor and EXor outburst
are not so evident in intensity, since both classes have
comparable outburst amplitudes, but in duration, remaining the
former bright for decades and the latter only for months. Indeed,
whether the EXor evolutionary stage follows the FUor one, or
merely represents a less evident manifestation attributable to the
same phase, is a matter not yet completely ascertained. As for the
FUor analogs, the EXor outbursts are believed to arise from
enhanced accretion events from a circumstellar disk, although
associated with lesser values of the mass accretion rate
($\dot{M}$). The quite uncertain picture which emerges in
interpreting the EXors phenomenum stems from the small number of
known objects and from the lack of a long-term multi-frequency
monitoring of the photometric and spectroscopic properties.

By examining the case of the eruptive object V1118 Ori, four
outbursts have been observed so far, although archival plates
suggest that there were further flares in the past (1939-1956 - Paul
et al. 1995). The first documented burst was during the period
1982-1984 (Chanal 1983; Parsamian \& Gasparian 1987); the second in
1988-1990 (Parsamian et al. 1993); the third in 1992-1994
(Garc\'{i}a Garc\'{i}a et al. 1995); the fourth in 1997-1998
(Hayakawa et al. 1998; Garc\'{i}a Garc\'{i}a \& Parsamian 2000). The
last outburst at the end of 2004 was announced by Williams \&
Heathcote (2005) and currently (September 2005), the object is
slowly fading but has not yet reached its quiescent phase.
Typically, the amplitude of the outburst (in V band) is larger than
3 magnitudes (from about 17-18 to 14.5) and the duration is longer
than 2 year, with a rise time of 8-10 months followed by a
long-lived decline phase of about 1.5 years. An optical spectrum
(from 3900 to 5600 \AA) was obtained (Gasparian et al. 1990) during
the rising period of the second outburst in the early 1989. It shows
emission lines of \ion{H}{i}, \ion{Ca}{ii}, \ion{Fe}{i} and
\ion{Fe}{ii} and the presence of P-Cyg profiles in the Balmer series
is suspected: this spectrum is similar to that of T Tauri stars with
moderate intensity. Prior to the most recent outburst, the source
was photometrically observed at near infrared wavelengths only three
times during both outburst and quiescence phases. We present here
the first near IR spectrum of V1118 Ori along with broad and narrow
band imaging data obtained during the last declining phase. A newly
obtained X-ray detection during the same phase is presented as a
result of a search in the existing archives. Accretion processes
have been suggested as the possible origin of the X-ray emission of
young low mass objects (e.g. Shu et al. 1997). However the
mechanisms are still unclear since many of these objects emit X-rays
via solar-like magnetically trapped coronal plasma (see the review
by Feigelson \& Montmerle 1999; Kastner et al. 2004a).

Archival X-ray data come from a multi-wavelength campaign to monitor
the 2004/2005 outburst of V1118 Ori. Such monitoring is conducted by
Audard et al. (2005), who have very recently presented their first
results. They focus mainly on the X-ray properties, although
accurately sampled optical and IR light curves are presented, as
well. They argue that the X-ray emission was due to a corona, but it
was influenced by the accretion during the outburst. Their time
coverage partially overlaps with our early observations which,
however, complement their results by covering a subsequent period,
relevant to ascertain the source behaviour.

The IR study carried out during the declining/quiescent phase allows
us to investigate how the properties of the circumstellar matter
prior the outburst will influence, through the accretion, the
outburst itself. The same study during the active phase samples how
the circumstellar material is altered by intermittent mass loss. The
X-ray investigation carried out during an optical/near IR monitoring
can help in discriminating between the mechanism(s) which
generate(s) the high energy emission in low mass protostars. Our aim
is manifold: (i) to better understand how the IR spectral signatures
are related to the intrinsic nature of an EXor; (ii) to derive a
consistent picture for both the IR and X-ray behaviour; and (iii) to
build up a systematic IR database useful as a reference for the next
monitoring of the quiescence and subsequent outburst phases. After a
short presentation in Sect.2 of our observations and data reduction
procedures, we provide and discuss the results in Sect.3, giving our
conclusions in Sect.4.

\section{Observations and data reduction}

\subsection{Near IR imaging and spectroscopy}

Near IR data were obtained at the 1.1m AZT-24 telescope located at
Campo Imperatore (L'Aquila - Italy) equipped with the
imager/spectrometer SWIRCAM (D'Alessio et al. 2000) which is based
on a 256$\times$256 HgCdTe PICNIC array. Photometry is performed
with broad band filters J (1.25 $\mu$m), H (1.65 $\mu$m) and K
(2.20 $\mu$m) along with a narrow band one, centered at 2.122
$\mu$m on the 1-0 S(1) H$_2$ transition. The total field of view
is 4.4$\times$4.4 arcmin$^2$ which corresponds to a plate scale of
1.04 arcsec/pixel. Low resolution ($\mathcal{R}$ $\sim$ 250)
spectroscopy is obtained by means of two IR grisms G$_{blue}$ and
G$_{red}$ covering the ZJ (0.83 - 1.34 $\mu$m) and HK (1.44 - 2.35
$\mu$m) bands, respectively, in two subsequent exposures. The long
slit is not orientable in position angle and samples a pre-defined
portion of the focal plane, 2$\times$260 arcsec along the
east-west direction. Details of the observations are given in
Table~\ref{journal:tab}. All the observations were obtained by
dithering the telescope around the pointed position. The raw
imaging data were reduced by using standard procedures for bad
pixel removal, flat fielding and sky subtraction. Continuum-free
images in the narrow band filter were obtained as a first step by
subtracting an appropriately scaled K image from the H$_2$ image.
Such scaling has been obtained by performing the photometry of a
number of stars located in different positions within the field.

Long slit spectroscopy was carried out in the standard
ABB$\arcmin$A$\arcmin$ mode with a total integration time of 800
sec. The observations were flat fielded, sky subtracted and
corrected for the optical distortion along both the spatial and
spectral direction. Telluric features were removed by dividing the
extracted spectra by that of a normalized telluric standard star,
once corrected for its intrinsic spectral features. Wavelength
calibration was derived from OH lines present in the raw spectral
images, while flux calibration was obtained from our photometric
data.


\begin{table*}
\caption[]{ Journal of observations
    \label{journal:tab}}
\begin{center}
\begin{tabular}{cccc}
\hline \hline\\[-5pt]
UT Date      & MJD$^a$ & Mode        & Filter/Grism      \\
\hline\\[-5pt]
Mar 20, 05   &  53449  &  Phot       & J,K               \\
Apr 03, 05   &  53463  &  Phot       & J,K               \\
Apr 15, 05   &  53475  &  Phot       & J,K               \\
Sep 06, 05   &  53619  &  Phot       & J,K, H$_2$        \\
Sep 11, 05   &  53624  &  Phot/Spec  & JHK, G$_b$,G$_r$  \\
\hline \hline
\end{tabular}
\end{center}
Notes to the Table:
\begin{itemize}
\item[a -] MJD = modified Julian Date.
\end{itemize}
\end{table*}


\subsection{X-ray observations}

Aiming to obtain reliable X-ray data on V1118 Ori, we searched the
public archives, finding the XMM-Newton results of pointed
observations taken during our monitoring period. The relevant
information on both the X-ray telescope and EPIC (PN and MOS)
cameras are given by Jansen et al. (2001), Str\"{u}der et al. (2001)
and Turner et al. (2001), respectively. Table~\ref{Xrayobs:tab}
summarizes the two XMM-Newton available observations. The first was
obtained few months after the most recent outburst (see also Audard
et al. 2005), while the second (on Sept. 08, 2005), obtained 7
months later, is almost simultaneous with our near IR spectrum,
taken 3 days after. The data have been processed using the
XMM-Newton Science Analysis Survey (SAS) v.6.1.0. We used the event
files linearized with a standard reduction pipeline (Pipeline
Processing System, PPS) at the Survey Science Center (SSC,
University of Leicester, UK).

Events spread at most in two contiguous pixels for PN (i.e.
pattern=0-4) and in four contiguous pixels for MOS (i.e,
pattern=0-12) have been selected.  Event files were cleaned from bad
pixels (hot pixels, events out of the field of view, etc.) and the
soft proton flares.

In order to remove periods of unwanted high background level, we
located the flares by analyzing the light curves of the count rate
at energies higher than 10 keV, where the X-ray sources
contribution is negligible. We rejected the time intervals, when
the count rate is higher than 11 (8 for 0212481101 observation)
counts s$^{-1}$ and 1.3 counts s$^{-1}$ for the PN and MOS cameras
respectively.  The source counts were extracted from a circular
region of 25 arcsec radius. The background counts were extracted
from the nearest source free region.  The response and ancillary
files were generated by the XMM-SAS tasks, {\sc rmfgen} and {\sc
arfgen}, respectively. The MOS1 and MOS2 spectra are combined
together. The spectra are binned to have at least 20 counts per
energy bin.

\begin{table*}
\caption[]{ X-ray detections parameters
    \label{Xrayobs:tab}}
\begin{center}
\begin{tabular}{ccccccc}
\hline \hline\\[-5pt]
obs.id.& UT Date & MJD$^a$ & EPIC & t$_{int}$ & \multicolumn{2}{c}{(counts s$^{-1}$)$^b$} \\
       &         &         & cam  &  (sec)    & (0.5-10 keV)  &  (0.5-2 keV)          \\
\hline\\[-5pt]
0212480301   &  Feb 18, 05  & 53419 & PN/MOS  & 1.7 10$^4$ &  0.0162  & 0.0132              \\
0212481101   &  Sep 08, 05  & 53621 &  PN     & 1.3 10$^4$ &  0.0044  & 0.0027               \\
\hline \hline
\end{tabular}
\end{center}

Notes to the Table:
\begin{itemize}
\item[a -] MJD = modified Julian Date.
\item[b -] counts are for the PN camera and are background subtracted.
\end{itemize}
\end{table*}

\section{Results and discussion}

\subsection{Near IR Imaging and Photometry}

\begin{table*}
\caption[]{ Near IR photometry of V1118 Ori
    \label{mag:tab}}
\begin{center}
\begin{tabular}{cccccc}
\hline \hline\\[-5pt]
MJD$^a$  & J & H  & K & J-H & H-K \\
\hline\\[-5pt]
48260-49350$^b$   &   12.21 (1) & 11.29 (1)  &  10.49 (1) & 0.92  & 0.80    \\
49426$^c$         &   8.52 (5)  & 8.04 (4)   &  7.94 (3)  & 0.48  & 0.10    \\
51872$^d$         &   12.64 (2) & 11.51 (3)  &  10.85 (2) & 1.13  & 0.66    \\
\hline\\[-5pt]
53449  &  11.16$^e$  &  -      &  9.79  & -     &    -    \\
53463  &  10.79      &  -      &  9.48  & -     &    -    \\
53475  &  10.94      &  -      &  9.53  & -     &    -    \\
53619  &  11.08      &  -      &  9.71  & -     &    -    \\
53624  &  11.23      &  10.45  &  9.85  & 0.78  & 0.60    \\
\hline \hline
\end{tabular}
\end{center}

Notes to the Table:
\begin{itemize}
\item[a -] MJD = modified Julian Date.
\item[b -] The authors (Hillenbrand et al. 1998) do not specify the
date of the observation which was taken during the period 1991-1993.
The errors (in units of 0.01 mag) are given in parentheses.
\item[c -] Data from Garc\'{i}a, Garc\'{i}a et al. (1995).
\item[d -] 2MASS observation.
\item[e -] Errors of our photometry in all the three bands never exceed 0.02
mag.
\end{itemize}
\end{table*}

\begin{figure}
 \centering
   \includegraphics [width=12 cm] {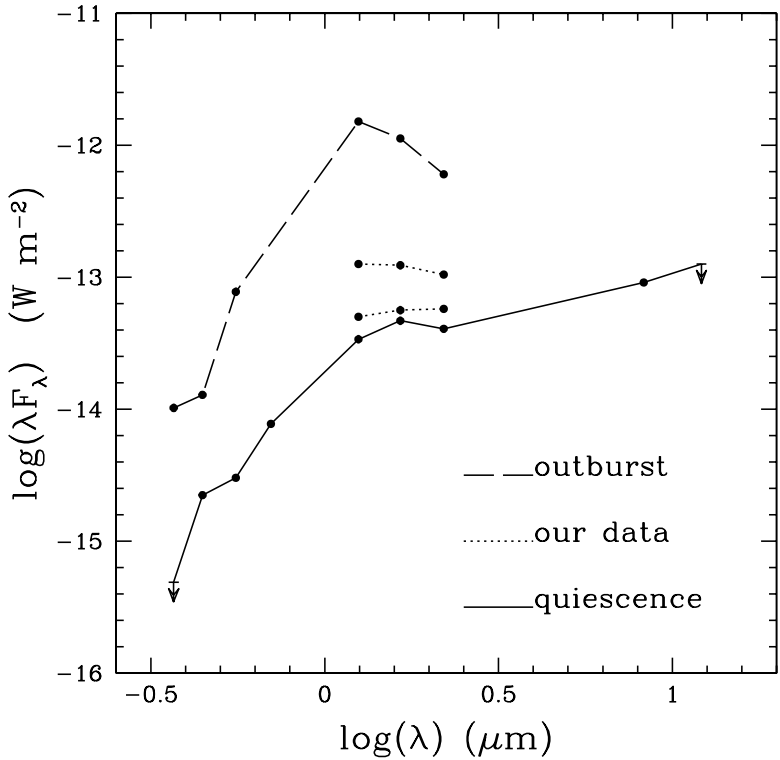}
   \caption{Observed SEDs of V1118 Ori constructed with literature and present paper IR data
   that correspond to different activity phases as indicated in the bottom right
   corner.
   \label{sed:fig}}
\end{figure}

In Table~\ref{mag:tab} the photometric data are presented.  For
the sake of completeness, we report in the first three lines of
Table~\ref{mag:tab} the near IR photometric data available in
literature prior to the recent outburst. The maximum brightness
corresponds to the third outburst active phase (Garc\'{i}a
Garc\'{i}a et al. 1995) and the minimum to the most recent
quiescent phase (2MASS All-Sky Catalog of Point Sources - Cutri et
al. 2003): the J,H and K brightness decreases by 4.1, 3.5 and 2.9
mag, respectively, thus less and less pronounced variations occur,
while the wavelength increases. The third available near IR
observation (Hillenbrand et al. 1998) was obtained during an
almost quiescent phase and is in agreement with the described
trend. Our results (from the fourth to the last line of
Table~\ref{mag:tab}), are sampling the slow declining after the
recent outburst and substantially agree with the results of Audard
et al. (2005). Since the data given in Table~\ref{mag:tab} refer
to different outburst and quiescence phases and are not taken
simultaneously with the optical data collected from the
literature, they cannot be used, in principle, to construct any
significant SED. However, we can assume that maxima and minima of
different epochs recur at similar flux levels, a circumstance
which is reasonably confirmed by the past record of optical
observations. Under this hypothesis we plot in
Figure~\ref{sed:fig} two SEDs, obtained by considering the weakest
and the strongest values, respectively. To enlarge as much as
possible the SED frequency range, we searched the IR archives
(IRAS, ISO, MSX) for a mid- far-infrared counterpart of V1118 Ori
and we were able to detect a faint emission (at a 2$\sigma$ level)
from a point-like source only in the MSX A-band at 8.28 $\mu$m
which corresponds to a flux density of 0.3 Jy, obtained during a
quiescent phase. This value suffers from a certain degree of
contamination, of about 30 \%, which is not due to the nearby
bright star V372 Ori, but to the irregular diffuse emission
characteristic of this sky area. In the MSX C-band (at 12.13
$\mu$m) only an upper limit of 0.5 Jy, corresponding to the
closeby sky emission, can be derived. This MSX determination,
although marginal, is quite important since permits both to
delineate the SED shape from the U band to the mid-IR and to
compute a reliable value of the bolometric luminosity (L$_{bol}$).
The MSX detection has been obtained during an inactive phase (July
1996), and, as such, it has to be attributed only to the quiescent
SED. We lack of mid-IR detection during an active phase, however
the source is expected to increase its brightness at all
wavelengths, as proven by Muzerolle et al. (2005). In fact, they
revealed the outburst of V1647 Ori (whose EXor or FUor nature is
still not ascertained) with {\it Spitzer Space Telescope},
reporting a brightening in all the 3.6-70 $\mu$m range. By
integrating the flux densities, we obtain L$_{bol}$ values of 1.4
(quiescence) and 25.4 (outburst) $L_{\sun}$, respectively. The
latter has to be considered as a lower limit, since in the
L$_{bol}$(outburst) computation we used the same flux at 8.28
$\mu$m ascribed to the inactive phase. These values indicate how
the increase in brightness is indeed impressive, a factor of more
than 18, and suggest that the V1118 Ori luminosity ($<$ 10
$L_{\sun}$) is more typical of the EXors than the 5-50 times more
luminous FUors.

The SEDs reported in Figure~\ref{sed:fig} present different shapes:
more peaked in the near infrared, looking like the T Tauri averaged
SED (D'Alessio et al. 1999), or rather flat which resembles that
typical of the eruptive objects (FUors-EXors) (K\'{o}sp\'{a}l et al.
2004, Muzerolle et al. 2005). All these shapes are characteristic of
accreting T Tauri stars  which show indeed variable SEDs in IR. The
apparent steepening of the SED during the outburst is due to a
single JHK photometry and, as such, does not necessarily apply to
all the outburst phases. Based on JHK data plotted by Audard et al.
(2005), a systematic variation of the SED slope with the activity
phase does not seem recognizable. A systematic IR monitoring of
repeated outbursts has to be accumulated to ascertain whether or not
the SED is flattening while the source is progressively fading.

\begin{figure}
 \centering
   \includegraphics [width=12 cm] {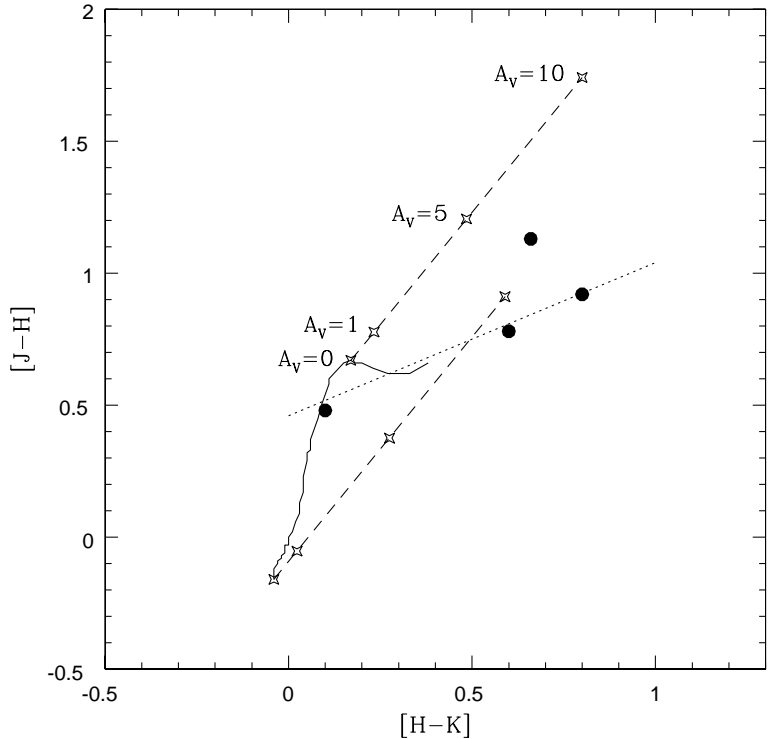}
   \caption{Locations of V1118 Ori on a near IR two colours diagram in different epochs.
   The solid line marks the unreddened Main Sequence, whereas
   the dotted one the locus pertaining to the T Tauri stars (Meyer et al. 1997).
   Dashed lines are the reddening law according to Rieke \&
   Lebofsky (1985); 10 mag intervals of A$_V$ are indicated by
   crosses.
   \label{colcol:fig}}
\end{figure}

The same aspect can be furtherly discussed by following as the
source behaves in the [J-H], [H-K] two colours diagram given in
Figure~\ref{colcol:fig}. The locations of V1118 Ori as detected in
epochs of different level of activity, indicate how the colours
are well in agreement with those predicted for T Tauri stars by
Meyer, Calvet \& Hillenbrand (1997). Such predictions result from
disk models with a range of accretion rates, inner disk radii and
viewing angles. We remark how the observed data are largely
consistent with unreddened values, implying that in the majority
of the cases the object is seen practically through a negligible
extinction (A$_V$ $\simeq$ 0). In one phase, far from the
outburst, it appears slightly redder and the same happens in the
optical band, as well. The very similar case of outburst recently
studied over a large range of frequencies, namely V1647 Ori,
behaves exactly in the same manner as V1118 Ori in the near IR
(Reipurth \& Aspin 2004; Vacca et al. 2004). With reference to
Figure~\ref{colcol:fig} the observed shift in colours is
compatible with an extinction value of A$_V$ $\sim$ 2 mag.
Variable scattering by some cavity walls might contribute in
making bluer the emerging light.

Several comparison stars are available within the 4$\arcmin$.4
SWIRCAM field; in particular one of them has remained stable
within 0.01 mag in JHK during our entire monitoring period, thus
it is used to construct the near IR light curves in terms of
differential photometry.
A slightly downward trend is superposed on statistically meaningful
variations (of about 0.2-0.3 mag) in the near IR on a shorter time
scale (days). Such rapid variations detected in all the JHK bands do
not seem related to extinction variability which should determine
larger peak to peak variations at shorter wavelengths. This
circumstance provides support to the idea that accretion processes
are dynamical on several timescales: non steady accretion and
rotational modulations occur from hours to weeks, while
instabilities of the magneto-spherical structure and EXors-FUors
events are characterized by longer timescales (months to years).

Our H$_2$ image at 2.122 $\mu$m shows no presence of any emission
down to a level of 1.1 10$^{-14}$ erg s$^{-1}$ cm$^{-2}$ (at
3$\sigma$) within a 4.4$\times$4.4 arcmin$^2$ FoV. The 1-0 S(1)
H$_2$ line at 2.122 $\mu$m is one of the best tracer of the gas
cooling after the passage of a shock front typically associated to
jet structures emerging from a central object. Our sensitivity in
revealing knots of H$_2$ emission (L$_{H_2}$ = 7 10$^{-5}$
L$_{\sun}$) would have allowed us to detect the faintest knots and
jets discovered by Stanke et al. (2002) during their unbiased
H$_2$ survey of Orion. No other IR spectral features indicative of
shock cooling (e.g. [\ion{Fe}{ii}] at 1.25 or 1.64 $\mu$m) are
detected in our spectrum (see below, Sect. 3.2). Although the
presence of an optical jet cannot be excluded, our observation
agrees with the lack of collimated jets. This is a quite common
feature in EXor and FUor objects that usually show powerful mass
loss only in form of isotropic winds.

\subsection{Near IR Spectroscopy}

\begin{figure}
 \centering
   \includegraphics [width=12 cm] {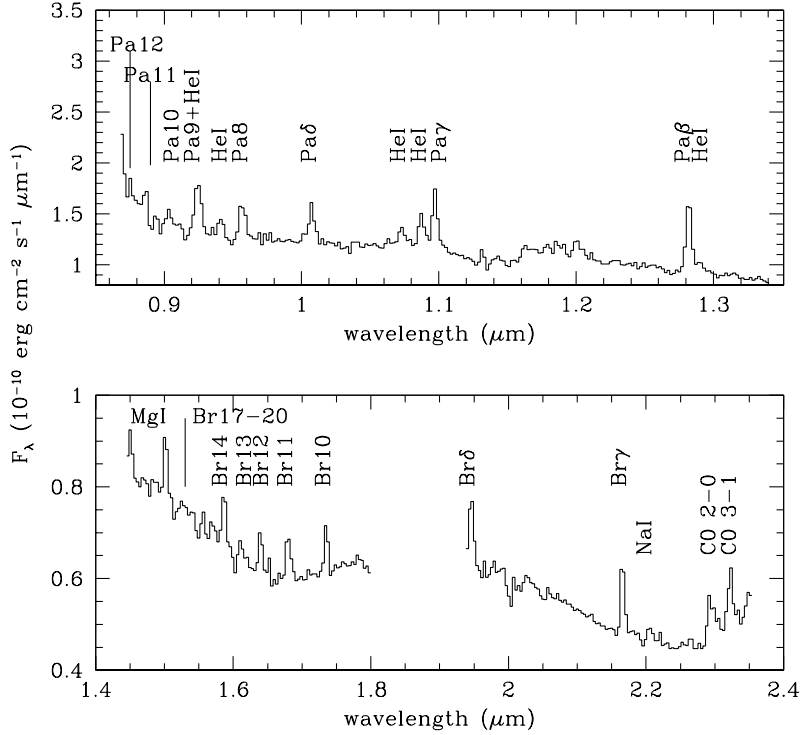}
   \caption{Near-IR spectrum of V1118 Ori. The lines Pa 11 and Pa 12 are identified here, but are
   neither listed in Table~\ref{lines:tab} nor used in our model, since their fluxes suffer large
   uncertainties due to the definition of a reliable local baseline.
   \label{IRspectrum:fig}}
\end{figure}

During the recent slowly declining phase we have obtained the
first near-IR spectrum of V1118 Ori depicted in
Figure~\ref{IRspectrum:fig} (see also Di Paola et al. 2005). The
derived line fluxes are given in Table~\ref{lines:tab}. It is an
emission line spectrum dominated by the hydrogen recombination
(Brackett and Paschen series) which signals the presence of
ionized gas close to the star. Our spectral resolution does not
allow to reveal whether lines have P Cygni profiles. We also
detect \ion{Mg}{i} line emission at 1.503 $\mu$m, weaker emission
features of \ion{He}{i} and a marginal (at S/N = 2) emission of
\ion{Na}{i} at 2.208 $\mu$m. These atomic features are commonly
found in the near-IR spectra of T Tauri stars and younger objects.
In particular, \ion{Mg}{i} and \ion{Na}{i} emission lines have
been recently revealed in another erupting object, namely V1647
Ori (Reipurth \& Aspin 2004; Vacca, Cushing \& Simon 2004), where,
however, \ion{He}{i} is present in absorption. We note that
\ion{Na}{i} line has been also detected, but in absorption,
(Herbig et al. 2001), in the near-IR spectrum of EX Lup, the
prototype of the EXor class. These circumstances suggest that an
IR spectroscopic data-base of eruptive variables, large enough to
cover different activity phases, has to be built up in order to
evaluate how possible similarities among EXors can be related to
the dominant mechanism(s) of gas excitation and cooling. Molecular
hydrogen contributions are absent and only the 1-0 S(1)
rovibrational line at 2.122 $\mu$m is marginally recognizable at a
1$\sigma$ level; such a spectroscopic circumstance confirms the
lack of any shock evidence also indicated by H$_2$ imaging
(Sect.3.1). The observed spectrum is much more similar to that of
an accreting T Tauri star (Greene \& Lada 1996) than the FUor
ones. All these latter, apart a couple of exceptions, have spectra
always dominated by absorption lines (Reipurth \& Aspin 1997).

We have checked that no contamination exists due to the HII region
associated to the nearby bright star V372 Ori. In
Table~\ref{lines:tab} the identified lines along with the measured
fluxes are given. The associated uncertainty refers to the rms of
the local baseline.

The CO overtone emission v=2-0, v=3-1 is clearly detected; the
same occurs in several young stellar objects (Carr 1989), although
in the majority of FUor CO bands are revealed in absorption
(Hartmann, Hinkle \& Calvet 2004). In all the cases CO emission is
seen together to Br$\gamma$ line emission, but the two features
likely come from different volume of gas. At temperatures of about
4000 K CO is dissociated and for values greater than 3000 K
molecular hydrogen is dissociated by collisions. However, for
density values greater than 10$^7$ cm$^{-3}$ and in presence of
H$_2$, CO dominates the cooling (Scoville et al. 1980). Therefore
the CO bands are specific probes of the circumstellar portions
where the gas is relatively warm at high densities. Carr (1989)
has quantitatively investigated two scenarios which account for
the observed CO emission: an accretion disk and a neutral stellar
wind with velocities of 100-300 km s$^{-1}$, where the gas is
assumed to be cool ($\sim$ 3000 K). In this latter case, which
agrees with our observations (see below this section), he provides
model predictions for deriving the mass loss rate from the CO
(2-0) luminosity (his Figure 8). The CO flux, calculated in the
2-0 band by integrating from 2.29 and 2.32 $\mu$m after continuum
subtraction, provides a CO luminosity L$_{CO}$ equal to 6.3
10$^{-5}$ L$_{\sun}$ and consequently a mass loss value in the
range (3-8) 10$^{-7}$ M$_{\sun}$ yr$^{-1}$ can be derived. The
uncertainty range accounts for the varying model parameters.

\begin{table*}
\caption[]{ Line emission fluxes of V1118 Ori
    \label{lines:tab}}
\begin{center}
\begin{tabular}{ccc}
\hline \hline\\[-5pt]
$\lambda_{vac}$ & Ident.& F $\pm$ $\Delta$F                   \\
($\mu$m)        &       & (10$^{-13}$ erg s$^{-1}$ cm$^{-2}$) \\
\hline\\[-5pt]
0.9017   &  Pa10        &  1.6 $\pm$ 0.1 \\
0.9230   &  \ion{He}{i} &  0.8 $\pm$ 0.1 \\
0.9231   &  Pa9         &  3.1 $\pm$ 0.1 \\
0.9548   &  Pa8         &  2.3 $\pm$ 0.2 \\
1.0052   &  Pa$\delta$  &  2.2 $\pm$ 0.2 \\
1.0670   &  \ion{He}{i} &  1.2 $\pm$ 0.3 \\
1.0832   &  \ion{He}{i} &  1.9 $\pm$ 0.2 \\
1.0941   &  Pa$\gamma$  &  2.5 $\pm$ 0.2 \\
1.2822   &  Pa$\beta$   &  4.1 $\pm$ 0.1 \\
1.2850   &  \ion{He}{i} &  1.0 $\pm$ 0.1 \\
1.5031   &  \ion{Mg}{i} &  0.8 $\pm$ 0.2 \\
1.5885   &  Br14        &  0.8 $\pm$ 0.2 \\
1.6114   &  Br13        &  0.5 $\pm$ 0.2 \\
1.6412   &  Br12        &  0.6 $\pm$ 0.2 \\
1.6811   &  Br11        &  0.9 $\pm$ 0.3 \\
1.7367   &  Br10        &  0.8 $\pm$ 0.2 \\
1.9451   &  Br$\delta$  &  1.0 $\pm$ 0.2 \\
2.1661   &  Br$\gamma$  &  1.2 $\pm$ 0.2 \\
2.2084   &  \ion{Na}{i} &  0.4 $\pm$ 0.2 \\
2.2935   &  CO 2-0      &  1.2 $\pm$ 0.4 \\
2.3227   &  CO 3-1      &  1.3 $\pm$ 0.3 \\
\hline \hline
\end{tabular}
\end{center}
\end{table*}

\begin{figure}
 \centering
   \includegraphics [width=12 cm] {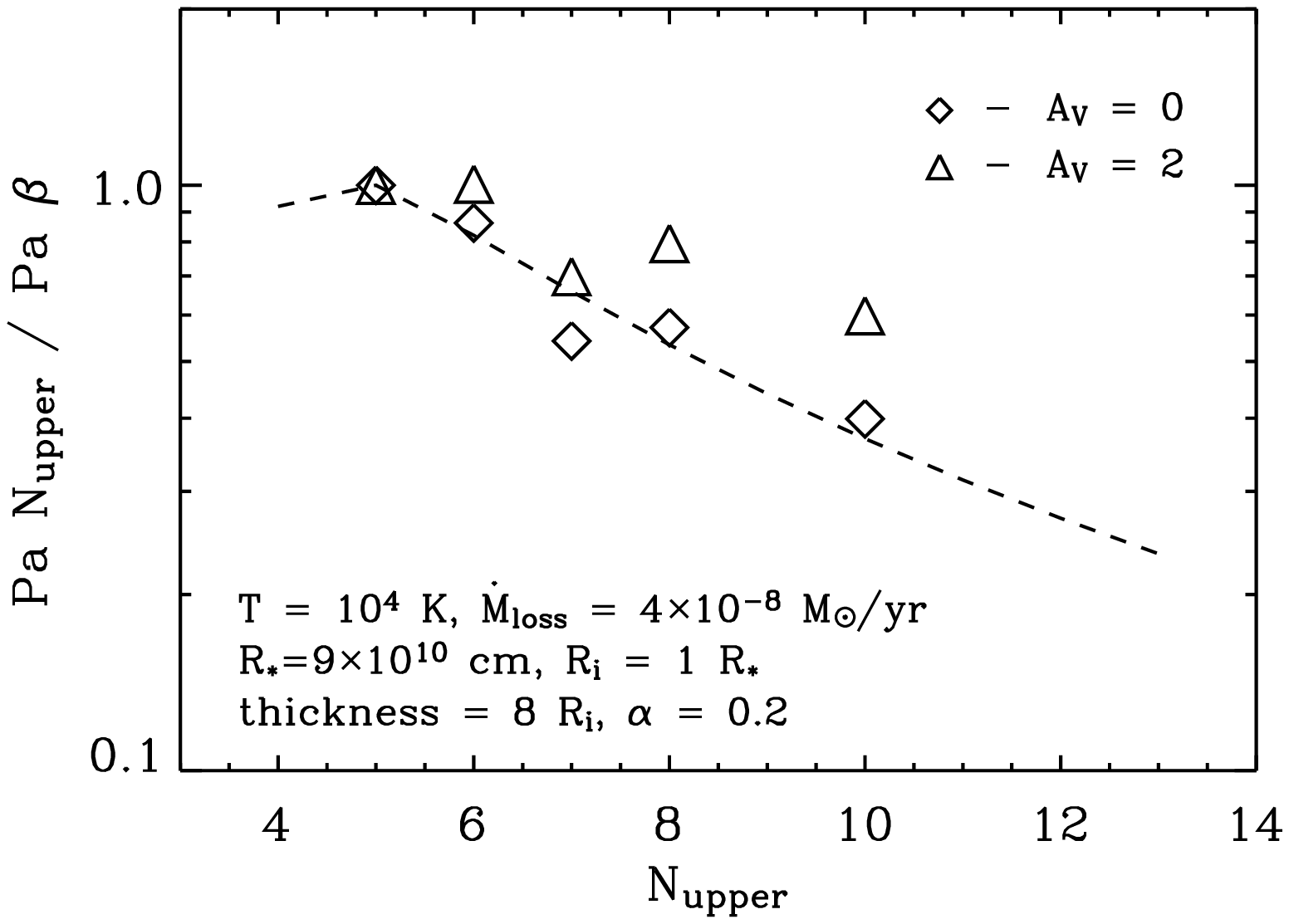}
   \includegraphics [width=12 cm] {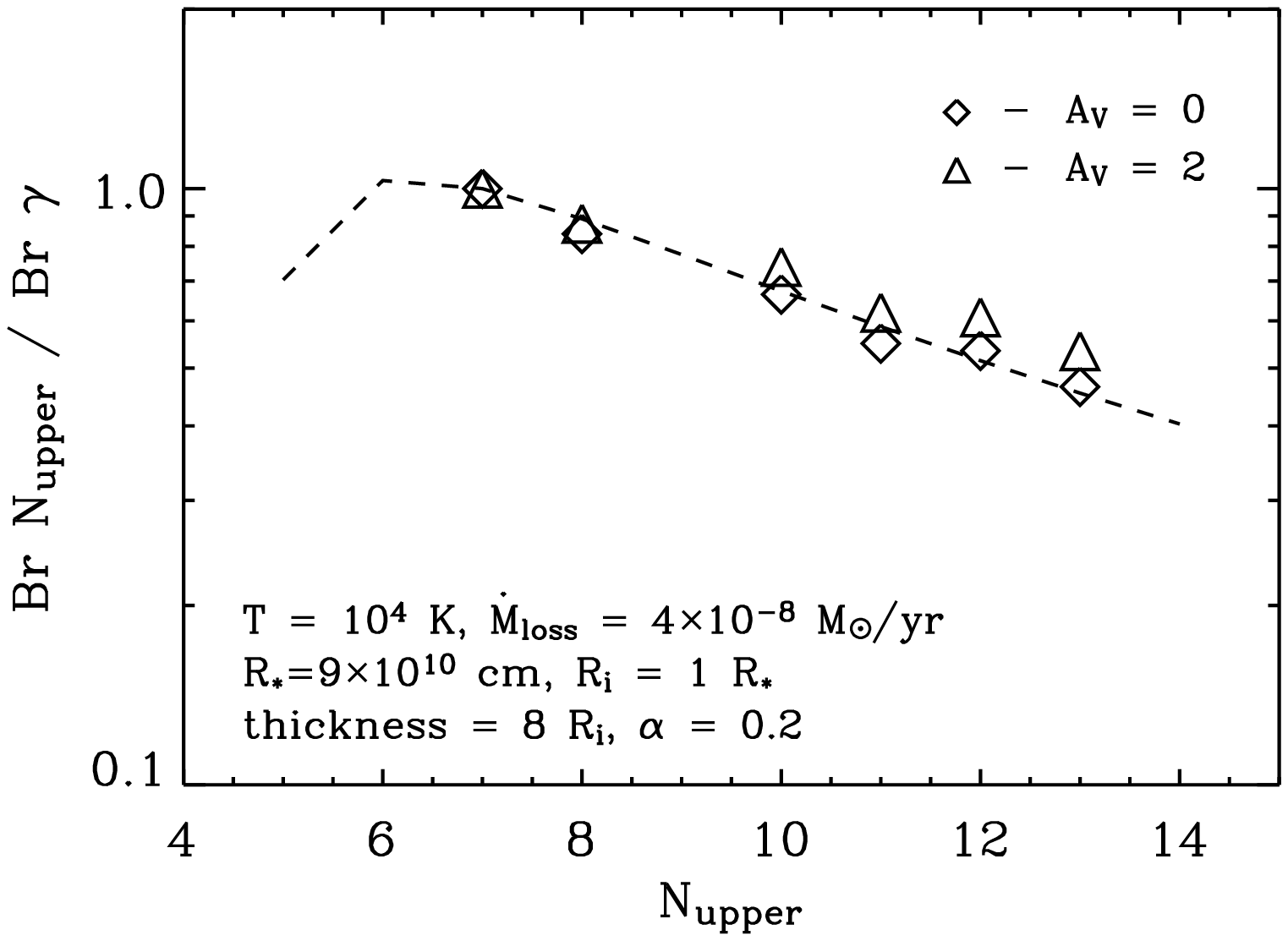}
   \caption{{\it Top Panel}: Line ratios of the Paschen series with respect to the Pa$\beta$ line.
   Diamonds and triangles refer to our observations undereddened (A$_V$ = 0) and extinction corrected
   (A$_V$ = 2), respectively. The dashed line represents the model line
   ratios. The relevant fit parameters are reported at the top of the panel. {\it Bottom Panel}: the same,
   but line ratios refer to the Brackett series with respect to the Br$\gamma$ line.
   \label{model:fig}}
\end{figure}

The observed \ion{H}{i} line emission has been compared with a wind
model (Nisini, Antoniucci \& Giannini 2004) which considers a
spherically symmetric and partially ionized envelope with a constant
rate of mass loss ($\dot{M}$ = 4$\pi r^2 \rho(r) v(r)$). The
emitting gas is assumed to be in LTE and the adopted gas velocity
law is:

\begin{equation}
{v(r) = {v_i + (v}_{max} - v_i)[1 - (R_{*}/R)^{\alpha}]}
\end{equation}

where $v_i$ and $v_{max}$ are the initial and maximum wind
velocity, respectively, while $R_{*}$ is the stellar radius. The
best fit to the data points is obtained for the following set of
parameters: gas temperature of 10$^4$ K, $v_i$ = 20 km s$^{-1}$,
$v_{max}$ = 200 km s$^{-1}$, envelope internal radius $R_{i}$ = 1
$R_{*}$, envelope thickness equal to 8 $R_{*}$, mass loss rate
$\dot{M}$ = 4 10$^{-8}$ M$_{\sun}$ yr$^{-1}$. An important
parameter of the model is the extinction value (A$_V$) for which
the line fluxes need to be corrected. From analyzing the position
of the object in the two colours  diagram (Sect.3.1), we have
assumed A$_V$ = 0, a value well consistent with our observations.
Figure~\ref{model:fig} shows the line ratios of the Brackett and
Paschen series with respect to the Br$\gamma$ and Pa$\beta$,
respectively, along with the best fit model, whose parameters are
indicated, as well. In addition, the model predicts for the
Br$\gamma$ a flux of 1.3 10$^{-13}$ erg s$^{-1}$ cm$^{-2}$,
practically equal to the observed one (1.2 10$^{-13}$ erg s$^{-1}$
cm$^{-2}$, see Table~\ref{lines:tab}). The effects of introducing
a different extinction value (A$_V$ = 2, namely the maximum
allowed by the observations) are also depicted in
Figure~\ref{model:fig}.

Finally, by ratioing the mass loss rates derived respectively from
the HI recombination (4 10$^{-8}$M$_{\sun}$ yr$^{-1}$) and the CO
emission ((3-8) 10$^{-7}$ M$_{\sun}$ yr$^{-1}$), a ionization
fraction of about 0.1-0.2 can be derived. This value is largely
consistent with those typical of active T Tauri environments
(Greene \& Lada 1996).

Aiming to evaluate the uncertainties on the derived parameters and
to check the sensitivity of our model, we have computed the range of
variation for each input parameter that is allowed to eventually
provide line flux predictions comparable (within a 50 \% extent)
with the observed values. Such analysis indicates that gas
temperature is one of the less sensitive parameters: variations
between 5000 K and 15000 K do not affect the fit significantly.
Conversely, other parameters are quite critical and their
variability ranges are consequently rather narrow; we obtain:
$\dot{M}$ = 4 $\pm$ 2 10$^{-8}$ M$_{\sun}$ yr$^{-1}$, R$_i$ = 1 - 2
R$_{*}$, thickness 8$\pm_{2}^{4}$ R$_{*}$. Finally we note that no
realistic solution can be found by using A$_V$ = 2 mag: indeed this
would imply an $\dot{M}$ increase of about 50 \% and a predicted
Br$\gamma$ flux exceeding the observed value by more than one order
of magnitude.

\subsection{X-ray properties}

\begin{figure}
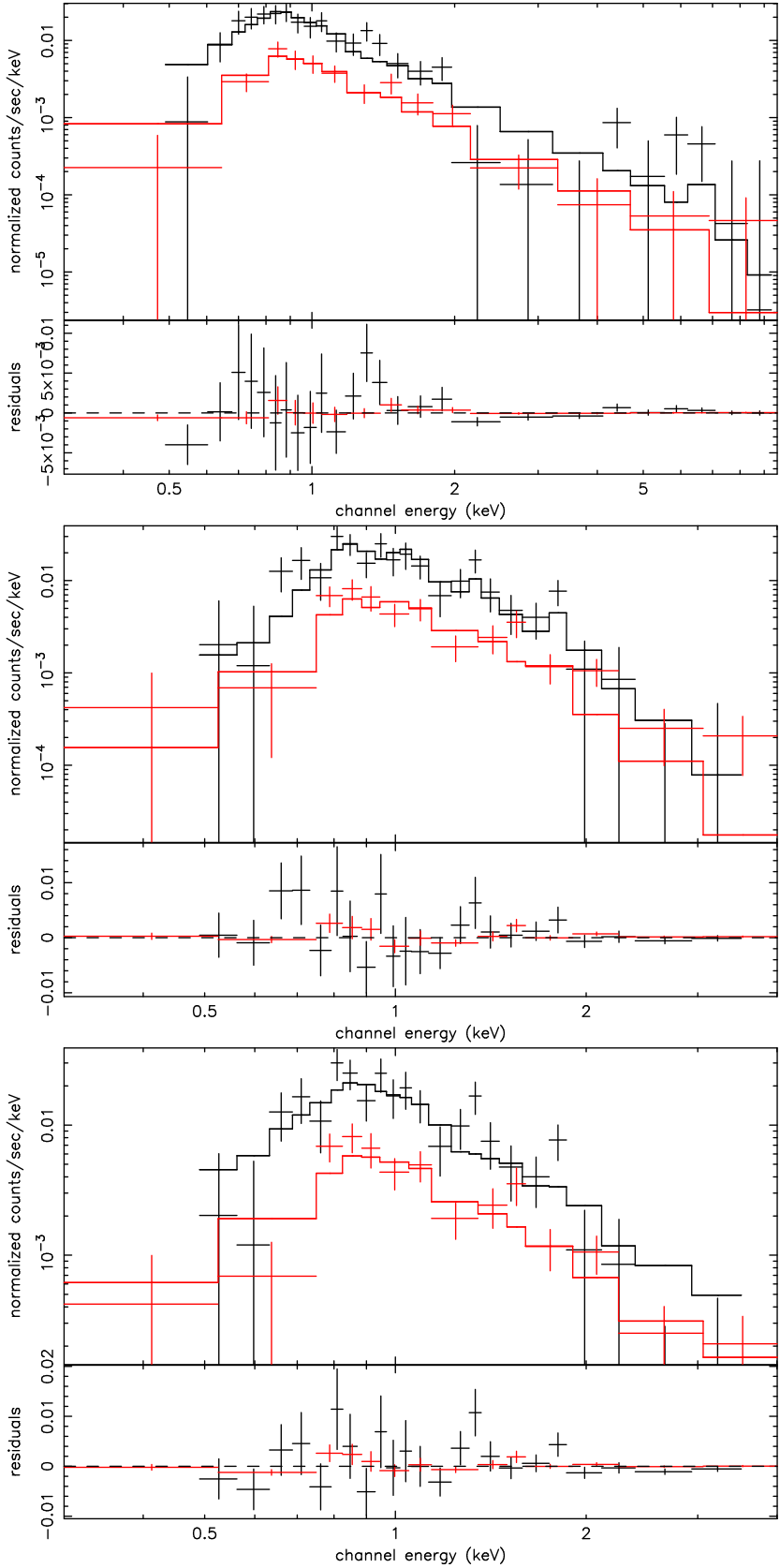

\centering
   \includegraphics [angle=-90,width=11 cm]{301_newplot.ps}

   \includegraphics [angle=-90,width=11 cm]{301_1mek_ab1_nhfree_sim4kev.ps}

   \includegraphics [angle=-90,width=11 cm]{301_2mek_ab1_nh019_sim4kev.ps}
   \caption{{\it Top panel:} X-ray spectrum (0.5 - 10 keV - February 2005) of V1118 Ori observed with the EPIC PN and MOS
   detectors on board XMM. A double-temperature thermal model has been adopted to fit the
   data (see text). The residuals computed in each bin are reported as well. {\it Middle panel:} as above,
   where the 0.5 - 3.5 keV portion is fitted with a single temperature model. {\it Bottom
   panel:} as above, where the 0.5 - 3.5 keV portion is fitted with a double temperature model.
   \label{Xspectrum:fig}}
\end{figure}

The XMM data were modeled using the XPEC spectral analysis
package. The first observation (see Table~\ref{Xrayobs:tab}), was
firstly fitted over the 0.5 to 10 keV range with an almost
unabsorbed two temperatures MEKAL thermal model (Mewe et al.
1995), which minimizes the $\chi^2$. The solar metal abundance has
been adopted for Orion (Esteban et al. 1998), however, a trial
with a value half than solar provided no significant improvement
in the $\chi^2$ value. In Figure~\ref{Xspectrum:fig} ({\it top
panel}) a comparison between the observed spectra with the best
fit model is given. The derived parameters, are given in the first
line of Table~\ref{Xraypar:tab}. By looking at the fit, the
suspicion could arise that a background contamination affects the
data at energies greater than 3-4 keV. Therefore, as a second
attempt, we fitted only the data points up to 3.5 keV. Following
the analysis done by Audard et al (2005), we fitted these (2005
February) data with a single temperature, obtaining the result
given in Figure~\ref{Xspectrum:fig} ({\it middle panel}), along
with the parameters reported in the second line of
Table~\ref{Xraypar:tab}; such parameters substantially agree with
the Audard et al. results. We also investigated the possibility to
fit the data with a two temperatures model obtaining the result
plotted in Figure~\ref{Xspectrum:fig} ({\it bottom panel}), and
the parameters given in the third line of Table~\ref{Xraypar:tab}.
The two temperature fit is as much acceptable as the single
temperature (see the residuals behaviour in
Figure~\ref{Xspectrum:fig} ({\it bottom panel})) and has a N$_H$
value which better agrees with a rather unobscured line of sight.
The circumstance that a good fit can be obtained with two
temperature values (about 30 and 7 MK), similar to those able to
account for the X-ray emission as detected prior the outburst
(Audard et al. 2005), confirms that X-ray emission of V1118 Ori is
practically unaffected by accretion events, being originated in
the corona. Thus, the eventual disappearance of a high temperature
component of the coronal emission, should occur with a certain
delay with respect to the outburst. The second spectrum reported
in Table~\ref{Xrayobs:tab} was obtained in Sept. 2005 (i.e. seven
months later). Although this latter X-ray spectrum is integrated
as long as the former, it appears intrinsically weaker and a
decent fit can be obtained only with a single temperature (of
about 10 MK) and by leaving the absorbing column density (N$_H$)
as a free parameter (fourth line of Table~\ref{Xraypar:tab}).
However, the best fit value, corresponding to a visible extinction
A$_V$ $<$ 2 (see Table~\ref{Xraypar:tab}) is consistent with that
of the earlier observation (A$_V$ = 1). This fact is quite
important since it proves that the X-ray fading is real and not
due to any intervening extinction effect. Remember that also the
near IR spectrum, taken in coincidental simultaneity (just 3 days
later), could be fitted only by assuming A$_V$ $\simeq$ 0. The
small difference between the extinction value derived from the
X-ray measurement by assuming a gas-to-dust ratio typical of the
interstellar medium and that derived from the simultaneously taken
near IR spectrum, confirms the suggestion that the gas-to-dust
ratio nearby the young stars could be larger than interstellar
(Kastner et al. 2004b). The Sept. 2005 observation corresponds to
L$_X$ (0.5-3.5 keV) = 0.6 10$^{-4}$ $L_{\sun}$ = 2.7 10$^{29}$ erg
s$^{-1}$; assuming that the true stellar luminosity ranges between
0.5 and 1 $L_{\sun}$, then Log(L$_X$/L$_{bol}$) $>$ -4.2, namely
the star is much more active than our Sun (for which
Log(L$_X$/L$_{bol}$) $\sim$ -6.5

By examining the data in Table~\ref{Xraypar:tab}, we remark: {\it
i}) a real fading of about a factor of 4 is occurred in the X-ray
emission (0.5 - 3.5 keV)  between February and September 2005;
{\it ii}) the existence of a temperature component, of about 7 MK,
during the active phases, interpreted as a coronal signature
(Audard et al. 2005) is a persistent feature even in the post
outburst phase, while the X-ray emission of the object begins to
decline. Recently, Preibisch et al. (2005), have presented their
survey (named COUP) of nearly 600 X-ray emitting T Tauri stars in
Orion and interpreted the widespread existence of a common
temperature around 10 MK, or less, as a real feature of the
coronal temperature distribution. Our data are fully in agreement
with their findings, confirming the active T Tauri nature of V1118
Ori. Attributing to the coronal activity the fairly constant
temperature component, detected even at a lower level of X-ray
emission, is in contrast with the interpretation given by Grosso
et al. (2005) of the recent eruption of V1647 Ori (Kastner et al.
2004b); however, the inconsistencies pertaining to the long
standing debate on the dominant X-ray mechanism in YSO, are
discussed by Audard et al. (2005).

The detected fading could be interpreted in the context of the
magnetically active stars that show a significant X-rays variability
(e.g. Feigelson \& Montmerle 1999), but its timing with an
optical-IR post-outburst phase, if not completely fortuitous, can
support a direct connection: a sort of modulation over the coronal
mechanism which tends to reduce the X-ray emission. Indeed,
according to Preibisch et al. (2005), the accretion events do reduce
the X-ray activity; we can only say that this effect cannot be
explained as due to intervening extinction, but to some other
effects (e.g. a weaker dynamo action related to the magnetic disk
locking).

\begin{table*}
\caption[]{ X-ray derived parameters
    \label{Xraypar:tab}}
\begin{center}
\begin{tabular}{cccccccccc}
\hline \hline\\[-5pt]
UT Date & $\chi^2$&d.o.f & N$_H$ &kT$_1$ & kT$_2$ & EM$_1$$^a$ &
EM$_2$$^a$& \multicolumn{2}{c}
{Flux (10$^{-14}$ erg s$^{-1}$ cm$^{-2}$)}\\
        &         &      & (10$^{22}$ cm$^{-2})$ &\multicolumn{2}{c}{(keV)} & \multicolumn{2}{c}{(10$^{52}$ cm$^{-3}$)} &
(0.5-10 keV)  &  (0.5-3.5 keV)     \\
\hline\\[-5pt]
Feb 18, 05& 35.7 & 35& 0.19$^b$       & 2.8$\pm_{1.1}^{3.8}$ & 0.6
$\pm$ 0.1 & 5 $\pm$ 1 & 2.5 $\pm$ 0.5  & 4.0 (5.8$^c$) & 2.4 (4.1$^c$)    \\
Feb 18, 05& 36.8 & 33& 0.8$\pm$0.1&          ---         &
0.52$\pm_{0.10}^{0.07}$ & ---
& 24 $\pm_{8}^{12}$ &---& 2.2 (23.7$^c$)\\
Feb 18, 05& 41.6 & 32& 0.19$^b$   & 2.5$\pm_{0.8}^{2.8}$ &
0.6$\pm$ 0.1 & 5 $\pm$ 1 &  2.0$\pm_{0.2}^{0.7}$
& --- & 2.2 (3.8$^c$)  \\
Sep 08, 05& 6.2  & 4 & $<$ 0.9$^d$& --- & 1.0$\pm_{0.6}^{1.1}$ & --- & $<$ 12$^e$ & ---  & 0.6 (1.1$^c$) \\
\hline \hline
\end{tabular}
\end{center}

Notes to the Table:
\begin{itemize}
\item[a -] The emission measure (EM) is derived, for each
temperature component, from the fit parameter K, through the
relationship EM = K$\times$10$^{14}$4$\pi$d$^2$ , where d is the
source's distance (in cm)
\item[b -] frozen parameter
\item[c -] unabsorbed flux
\item[d -] the best fit value corresponds to N$_H$ = 0.4 10$^{22}$
cm$^{-2}$
\item[e -] the best fit value corresponds to EM = 2.5 10$^{52}$
cm$^{-3}$
\item[-] the quoted errors correspond to a 90\% confidence level
\end{itemize}
\end{table*}

\section{The emerging picture and concluding remarks}

In this section we summarize the results presented above trying to
delineate a coherent picture for V1118 Ori. Firstly, in
Table~\ref{parameters:tab} the relevant information derived in the
present work is listed along few literature data.


\begin{table*}
\caption[]{V1118 Ori derived parameters.
    \label{parameters:tab}}
\begin{center}
\begin{tabular}{ccc}
\hline \hline\\[-5pt]
\multicolumn{2}{c}{Parameter$^a$} &  Value  \\
\hline\\[-5pt]
Distance                &   d          &    460 pc                            \\
Spectral Type           &              &    M1e                               \\
Stellar radius          & R$_{*}$      &    1.29 R$_{\sun}$                   \\
\hline
Bolometric luminosity   & L$_{bol}$    &  1.4 L$_{\sun}$ (min)                \\
                        &              &  $>$ 25.4 L$_{\sun}$ (max)           \\
X-ray luminosity        & L$_{X}$(0.5-3.5 keV)  &  0.6-2.2 10$^{-4}$ L$_{\sun}$ \\
Plasma temperature      & T            &  7 - 11 10$^6$ K                     \\
Visual extinction       & A$_{V}$      &  0 - 2 mag                           \\
Wind velocity           & v$_{w}$      &  200 km s$^{-1}$                     \\
Mass loss rate          & $\dot{M}$    &  4 $\pm$ 2 10$^{-8}$ M$_{\sun}$ yr$^{-1}$ (ionized)    \\
                        &              &  3-8 10$^{-7}$ M$_{\sun}$ yr$^{-1}$ (neutral)  \\
Ionization fraction     &              &  0.1 - 0.2                           \\
Envelope size           &              &  8$\pm_{2}^{4}$ R$_{*}$       \\
\hline \hline
\end{tabular}
\end{center}
Note to the Table:
\begin{itemize}
\item[a -] The first three lines of the table list literature
parameters, while the remaining are derived in the present work.
\end{itemize}
\end{table*}


V1118 Ori appears to be an accreting young object which undergoes
quasi periodic outbursts whose monitoring, so far conducted only in
the visual band has been started also in the near IR and X-ray
bands. The first marginal detection in the mid-IR (near 8 $\mu$m) is
also provided. Our observations largely confirm its EXor nature: its
bolometric luminosity increases from 1.4 $L_{\sun}$ to more than 25
$L_{\sun}$ passing from inactive to active phases.

V1118 Ori presents a SED typical of an accreting T Tauri star and
also its near-IR spectrum strictly resembles that of T Tauri,
showing strong \ion{H}{i}, \ion{He}{i} and CO features in emission.
The near-IR colors do not seem to undergo extinction variations and
are always compatible with low A$_V$ values (0-2 mag). Beside the
accretion disk, the star has a quite compact and partially ionized
circumstellar envelope created by the mass loss in stellar winds
which are likely accretion-generated. V1118 Ori is presumably in a
late stage of its pre-main sequence evolution and this is likely the
reason of its unobscured appearance. The lack of any collimated
outflow driven by the source provides further support to this
hypothesis. Large part of the relevant physics of V1118 Ori
originates in its accretion disk. By studying its X-ray properties,
a coronal origin is confirmed as the responsible for the high energy
emission with some influence by accretion. The parameters derived by
X-ray observations (plasma temperature and ratio L$_X$/L$_{bol}$)
are all in agreement with those obtained for active T Tauri stars.

All the observational evidences presented in this paper show the
similarity of V1118 Ori with low luminosity unobscured T Tauri
stars: this behaviour suggests that: (i) the EXor evolutionary
stage appears to follow the FUor one, more likely than to be a
less evident manifestation of the same phase; and (ii) the scanty
number of known EXor could be simply due to the difficulty of
performing an accurate multi-frequency monitoring of all the
active T Tauri stars.

We will continue our IR monitoring program by extending the
investigation also to shorter timescales (hours, days). Indeed
significant clues of rapid variations are already presented here
(Sect. 3.1), but wide databases have to be accumulated to properly
study to what extent the accretion processes are dynamical on
several timescales.

\begin{acknowledgements}
The authors would like to thank the referee Marc Audard for his
constructive comments on different aspects of this paper.
\end{acknowledgements}

\end{document}